# Crab: the standard X-ray candle with all (modern) X-ray satellites

M.G.F. Kirsch[1a], U.G. Briel[b], D. Burrows[c], S. Campana[d], G. Cusumano[e], K. Ebisawa[f], M.J. Freyberg[b], M. Guainazzi[a], F. Haberl[b], K. Jahoda[g], J. Kaastra[n], P. Kretschmar[a], S. Larsson[h], P. Lubinski[i], K. Mori[j], P. Plucinsky[k], A.M.T. Pollock[a], R. Rothschild[l], S. Sembay[m], J. Wilms[o], M. Yamamoto[j]

[a]European Space Agency, ESAC, Apartado 50727, 28080 Madrid, Spain, [b]MPE, Garching, Germany, [c]Penn State University, USA, [d]Osservatorio Astronomico di Brera, Italy, [e]IFCAI/CNR, Italy, [f]Astro-E2 Guest Observer Facility, USA, [g]NASA's GSFC, USA, [h]Stockholm Observatory, Sweden, [i]CAMK, Warsaw, Poland; INTEGRAL Science Data Center, Switzerland, [j]University of Miyazaki, Japan, [k]Harvard-Smithsonian Center for Astrophysics, Cambridge, USA, [l]Center for Astrophysics & Space Sciences, University of California, San Diego, USA, [m]Dept. of Physics and Astronomy, Leicester University, Leicester, UK, [n]SRON, The Netherlands, [o]University of Warwick, Coventry, UK

## ABSTRACT

Various X-ray satellites have used the Crab as a standard candle to perform their calibrations in the past. The calibration of XMM-Newton, however, is independent of the Crab nebula, because this object has not been used to adjust spectral calibration issues. In 2004 a number of special observations were performed to measure the spectral parameters and the absolute flux of the Crab with XMM-Newton's EPIC-pn CCD camera. We describe the results of the campaign in detail and compare them with data of four current missions (Integral, Swift, Chandra, RXTE) and numerous previous missions (ROSAT, EXOSAT, Beppo-SAX, ASCA, Ginga, Einstein, Mir-HEXE).

## 1. INTRODUCTION

Since the early days of X-ray astronomy the Crab nebula has been used as a calibration target for balloon, rocket and satellite-borne X-ray detectors. The major X-ray missions still operating in orbit are (in order of launch date): RXTE, Chandra, XMM-Newton, Integral and Swift. We shall describe the cross-calibration of these operational observatories and numerous previous experiments.

The Crab was optically discovered in 1054 by Chinese astronomers who observed a supernova event as a so-called bright guest star that was observable in the sky with the naked eye for some days during the day and for the following two years at night. In 1731 the source was then observed by John Bevis, a British physicist and amateur astronomer and was first catalogued in 1758 by Charles Messier as M1 in his eponymous catalogue. The interest in the Crab increased in 1942 when Baade[1] performed detailed observations of the structure of the nebula.

In the X-ray regime in 1963 Gursky et al.[2] discovered a source which was shortly afterwards localised by Bowyer et al.[3] in a 2°x 2° error box in the Crab region. The Crab can be localised in the optical in an elliptical region of 180"x120" while in the X-ray regime it is slightly smaller. Figure 1 shows overlays of the Crab nebula in various energy ranges. Staelin & Reifenstein[4] discovered the radio pulsar in the Crab nebula in 1968.

In the UHURU catalogue the Crab is listed as the fifth strongest X-ray source[5]. Toor and Seward[6] argued that the Crab can be interpreted as a constant intensity source, since most of the emission is from the diffuse part of the nebula, which is extended over a diameter of four light years. Therefore the nebula flux should not vary on time scales shorter than a few years. The pulsar flux however is a possible source for variation. The pulsar is believed to inject high-energy electrons into the nebula that radiate via synchrotron emission. If the engine of production of those electrons were variable one would expect a change of the Crab spectrum. This behaviour is not observed. Toor and Seward used a simple power law spectrum of the form $I = N(h\nu)^{-\alpha}$ (where $I$ is the differential energy flux and $h\nu$ the photon energy) to fit the Crab spectra containing the diffuse and pulsed emission. This was done for various observations performed with balloon, rocket and satellite-borne proportional Geiger and scintillation counters in energy ranges from 0.2-500 keV. They concluded that a joint fit to past data was a power law with the parameters α=1.08±0.05 and $N$=9.5 photons keV$^{-1}$cm$^{-2}$s$^{-1}$ at 1 keV and the best spectral shape between 2 and 50 keV measured from their own experiment was a power law with the parameters α=1.1±0.03 and $N$=9.7±1.0. Note that in modern *XSPEC* notation a power law is described as $I/(h\nu) = N(h\nu)^{-\Gamma}$ where $\Gamma = \alpha + 1$. These values for α (Γ) and $N$ have subsequently been generally adopted as 'canonical' parameters of the Crab spectrum in the X-ray range, with the caveat that some authors have used the fit to the past data (α=1.08, $N$=9.5), while others the values of Toor and Seward (α=1.10, $N$= 9.7). Observations in the gamma-ray range (e.g. Kuiper et al.[24]) typically indicate a steeper spectral shape but there is no general agreement on the best overall spectral model or the nature of the Crab steepening.

---

[1] mkirsch@sciops.esa.int +34 91 8131 345; fax +34 91 8131 172

Contrary to the optical regime the X-ray regime does not have a set of so-called "standard stars" that can easily be used for calibration, because the X-ray sky is much more variable than the optical. Furthermore, experience has shown that it is difficult to first calibrate instruments on the ground and then maintain this calibration through launch and the harsh conditions in space that lead often to change of detector behaviour with operation time in orbit. Especially for absolute calibration, in principle only clusters of galaxies and supernova remnants should be used because they are supposed not to be variable on human-related time scales. Nørgaard-Nielsen et al.[7] proposed in 1994 a mission called eXCALIBur to establish a standard set of X-ray calibration sources including the Crab. But no further effort was put into that direction. However, from the late 1970s on, the Crab was used as a de facto standard X-ray calibration source for lots of missions because its spectrum is simple and - at least the nebula region - not variable. Many instruments have been adjusted in their in-orbit calibration to fit the standard spectral Crab parameters even in contradiction, sometimes, to ground calibrations. This paper tries to give a clear view of which instruments have been adjusted in orbit through their calibration to the canonical Crab parameters and which have not. Furthermore we will give the results of spectral parameters for all instruments and discuss where and why they may differ from the canonical values and whether the non "Crab fudged" instruments may be closer to the truth.

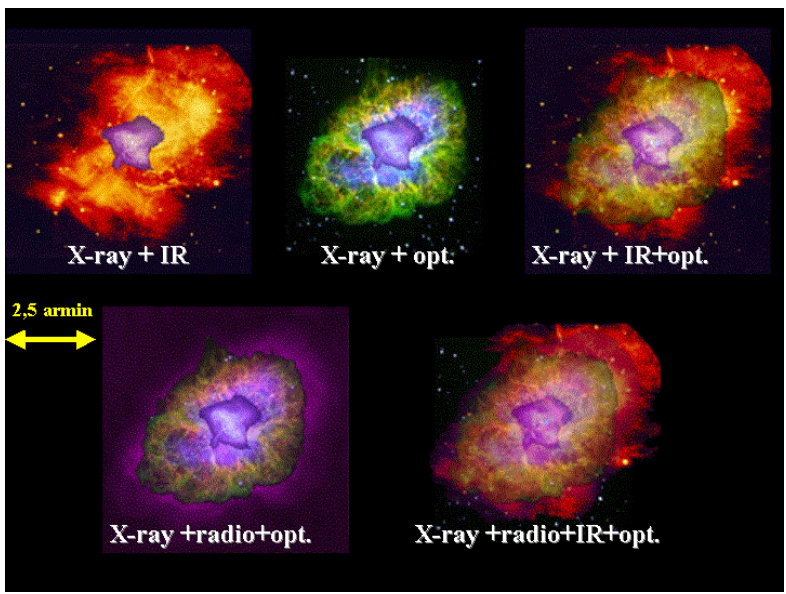

**Figure 1: The Crab nebula in different energy regimes. (Courtesy of Chandra, Harvard web pages)**

## 2. THE BEST-FIT MODEL FOR THE CRAB

As mentioned in the introduction the Crab spectrum can be fitted with a simple absorbed power law. In recent years various authors have discussed the question on the correct absorption model. Based on an updated compilation of UV and optically derived abundances, Wilms et al.[8] proposed new abundances with regards to the default XSPEC abundances of Anders and Grevesse[46]. For the Crab direction under-abundances in oxygen, neon and iron have been claimed by Willingale[9] and Weisskopf [10]. Weisskopf et al. found that the abundances towards the Crab are much closer to the Wilms et al. abundances than to others, which is the motivation for using them in this paper. Knight[11] and Jung[12] measured the pulsed and nebular spectral components of the Crab with HEAO-1 in the 18-180 keV range. The former found the phase-averaged, pulsed power law index to be 1.83± 0.03 while the latter found the nebular component power law index to be 2.13±0.05. Willingale[9] used the imaging properties of XMM-Newton to separate the two components, and found the pulsed index to be 1.63±0.09 and the nebula average index to be 2.108±0.006. One must keep these two components in mind when calibrating X-ray and gamma ray instruments. However after various checks we decided to use for this analysis of individual spectra the XSPEC model: *phabs\*powerlaw* using abundances of Wilms et al. and cross-sections of Balucinska-Church & McCammon or Verner et al.. This is accurate enough for the energy range of a single instrument. For the joint fits we use *const\*phabs\*powerlaw* freezing the EPIC-pn constant to 1 and linking all parameters of the other instruments to the EPIC-pn parameters allowing only the constants of the individual instruments to vary. This will provide new canonical Crab parameters and show the deviation from every single instrument as a function of energy. We also tried a double power law for the analysis (especially for the RXTE) though, for the overall analysis, this did not improve the fits and showed no difference to the general picture.

## 3. RESULTS OF XMM-NEWTON

XMM-Newton[13] has been operating since December 1999 with six instruments in parallel on its 48-hour highly elliptical orbit. Three Wolter type 1 telescopes with 58 nested mirror shells focus X-ray photons on the five X-ray instruments of the European Photon Imaging Camera (EPIC)[14,15] and the Reflecting Grating Spectrometers (RGS)[16]. The Optical Monitor (OM)[17] using a 30 cm Ritchey Chrétien optical telescope can perform parallel optical observations of the same field. EPIC consists of two parts: EPIC-MOS (Metal-Oxide Semi-conductor) and EPIC-pn (p-n-junction). The two EPIC-MOS cameras use front-illuminated MOS-CCDs as X-ray detectors while the EPIC-pn camera is equipped with a pn-CCD, which has been specially developed for XMM-Newton. EPIC provides spatially resolved spectroscopy over a field-of-view of 30' with moderate energy resolution. The EPIC cameras can be operated in different observation modes related to the different readouts in each mode[18, 19, 20, 21]. The RGS is designed for high-resolution spectroscopy of bright sources in the energy range from 0.3 to 2.1 keV. The OM extends the spectral coverage of XMM-Newton into the UV and optical, and thus opens the possibility of testing models against data over a broad energy band. Six filters allow colour discrimination, and there are two grisms, one in the UV and one in the optical, to provide low-resolution spectroscopy.

The following sections describe the kind of data reduction for the individual instruments. Results for the standard *phabs\*powerlaw* fit are given and summarized in Tab. 2, with results of other models where appropriate in the following subsections.

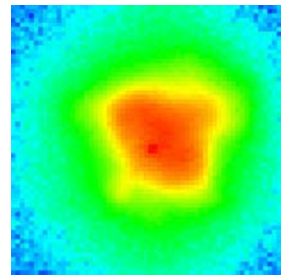

Figure 2: EPIC-pn Crab observation in SW mode

### 3.1. EPIC-pn

The EPIC-pn camera, which provides XMM-Newton's highest time resolution (Timing Mode: 30 μs, Burst Mode: 7 μs) and moderate energy resolution (E/dE = 10 to 50) in the energy band from 0.2 to 15 keV is the ideal instrument for observations of bright sources like the Crab. In the Burst mode the pile-up limit for a point source is 60000 cts/s, which corresponds to a maximum flux of 6.3 Crab. However, the special readout in the Burst mode leads to a loss of spatial resolution in the shift-direction. Moreover, the Burst mode livetime is only 3 %, making it useful only for a limited type of observations of bright sources like the Crab.

The Burst mode operates a special readout similar to a tape recorder in that 200 lines are fast-shifted within 14.4 μs while accumulating information from the source. Like in Timing mode, this leads to a loss of spatial resolution in the shift-direction. The stored information is then read out as normal, where the last 20 lines have to be deleted because of contamination by the source during the readout. The CCD is then erased by a fast shift of 200 lines, immediately after which the next Burst readout cycle starts. In the following we use *xmmsas* terminology for the images, in which RAWY represent the position in the CCD in shift direction from 1-200 and RAWX perpendicular to the shift direction from 1-64. Note that all pixels with RAWX=35, for example, are also called "column 35".

The Crab is used as a calibration target for XMM-Newton's timing, with observations performed on a routine basis every year. However, since these observations are optimised for timing analysis and performed in pn-Timing and pn-Burst mode, XMM-Newton points to the pulsar position of the nebula. In this configuration parts of the nebula are not imaged in the active CCD and can therefore not be used for absolute spectrophotometry. In 2004 we performed 3 observations of the Crab, where the pointing was optimised such that the full nebula fell on the active CCD areas. This enables us to measure the correct spectral shape and flux of the Crab nebula. We verified the correct pointing with a SW mode observation in the same revolution (0160960501) that shows in Fig. 2 that the Crab was fully seen by CCD 4.

The data have been extracted from the observation 0160960401 in revolution 874. EPIC-pn was operated in its special Burst mode. In addition we had to set not only the usual row 180-199 to bad, but also 12 additional rows in order to avoid telemetry problems and pile-up from that region during the slow read out of the CCD after the fast shift in Burst mode. In Burst mode the spatial information is lost and therefore this judgment cannot be made from the actual observation used for the spectral analysis. In order to avoid telemetry problems we switched off both MOS cameras and gave full telemetry to the EPIC-pn. The effective exposure time of the Burst observation was 437 sec. This corresponds to an observation time of around 14.5 ksec because the effective exposure of 3 % in Burst mode mentioned above. The data sets were processed using *SAS v6.1* with a special setting of *epchain* that enables the software to correct for the different source location in the detector resulting in a different Charge Transfer Inefficiency (CTI) correction. *(epchain datamode=BURST withsrccoords=yes srcra=83.633208 srcdec=22.014194)* Note that this processing is currently not possible with *epproc* but will be enabled in *SAS v6.5*.

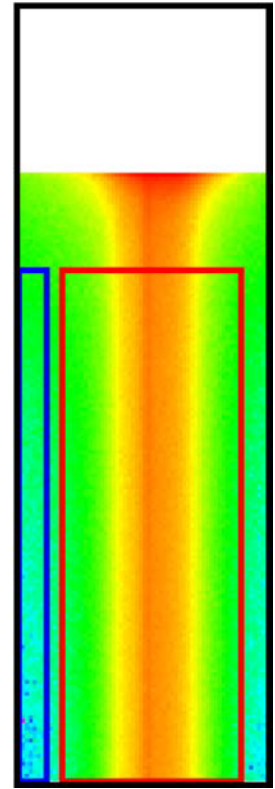

**Figure 3: Crab image in Burst mode. The solid black box indicates the CCD area, the red box the extraction region for the source spectrum and the blue box the extraction region for the background spectrum.**

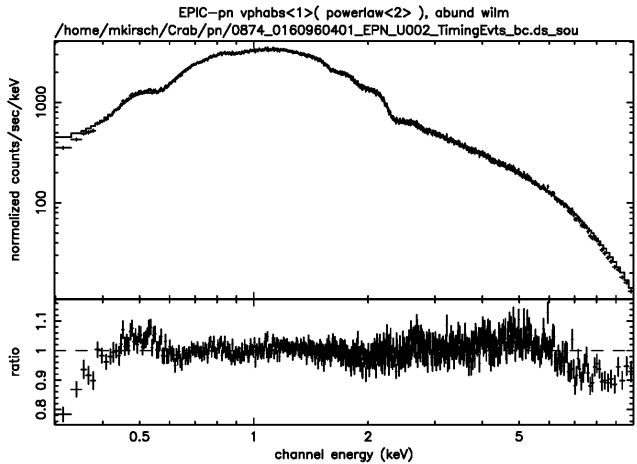

**Figure 4: Spectrum of the Crab nebula from EPIC-pn using a *vphabs\*powerlaw* fit. The residuals around 2.2 keV and 5 keV may be due to uncertainties in the effective area calibration of the XMM mirrors. The flux deficit at high energies is also believed to be a calibration uncertainty.**

We generated spectra for the source region *((RAWX,RAWY) IN box(34.75,67.5,23.75,65.5,0))* and background region *((RAWX,RAWY) IN box(4.5,67,3.5,65.5,0))* using single and double events. Figure 3 shows the extraction regions. Note that not all rows can be used in Burst mode in order to avoid contamination of the spectra by piled-up events that happen to fall during the slow readout in some of the high RAWY rows seen also in Fig. 3 as the halo-like structure in the upper part of the image. We generated standard response matrices and ARF files with the *SAS* and used *XSPEC* to fit the data in the 0.3-10 keV energy range with two models: 1) *phabs\*powerlaw;* and 2) *vphabs\*powerlaw* in which abundances for O, Ne and Fe were allowed to vary. We derived for 1) the values in Table 2 and for 2) $\Gamma=2.118\pm0.005$ with a normalisation of $8.56\pm0.05$ and $N_H= (4.04\pm0.03)\cdot10^{21}$ cm$^{-2}$. For the variable abundances we found O: $0.93\pm0.02$, Ne: $0.86\pm0.09$ and Fe: $0.76\pm0.08$. Figure 4 shows the EPIC-pn spectrum.

## 3.2. EPIC-MOS

The MOS data are from XMM-Newtons revolution 698. We analysed only the MOS2 data in Refresh Frame Store mode. In order to avoid pile-up we used only the out-of-time events of the whole Crab. This unfortunately does not give a reliable normalization, although it does allow the photon index and $N_H$ to be determined. Note that for the analysis a new RMF has been used taking the MOS time evolution into account. These RMFs will be generally available with the next SAS release.

## 3.3. RGS

Thanks to its high spectral resolution, the depths of the absorption edges in the Crab spectrum due to interstellar gas can be measured much better with the RGS than with any other instrument, leading to accurate column densities of individual elements. Nonetheless, some caution is necessary because the nebula has a significant dust-scattering halo[22] which may scatter a wavelength-dependent fraction outside the RGS aperture that could be serious at the longest wavelengths. Using Rosat PSPC data, Predehl and Schmitt[22] estimated that the halo contains 9.1 % of the total flux at a mean photon energy of 1.02 keV.

However, as the scattering fraction is a smooth function of energy, we proceeded as follows. The Crab nebula spectrum was extracted from data in which only 3 CCDs were used in order to avoid pile-up, which amounts to a few percent though with no expected effect on the absorption edges. Data were extracted using *SAS v6.1* and incorporated the latest effective area corrections before model fitting using the *SPEX* package; for the continuum emission we took a model derived by Kuiper et al.[24] based on BeppoSAX and other higher-energy instruments. Superimposed on this, we used the *hot* model of *SPEX* to model the transmission by the interstellar neutral gas. This model incorporates continuum and line absorption from a gas in collisional ionisation equilibrium. We fixed the temperature of this gas to 0.5 eV, essentially making the plasma neutral. The absorbed continuum was then convolved with the spatial profile of the Crab nebula in the dispersion direction, simply approximated by a Gaussian of adjustable width and centroid. Abundances and the hydrogen column of the absorber were free parameters, as well as the normalisation, but not the shape, of the continuum. Reference abundances were taken from the recent protosolar compilation of Lodders[25]. The results of our fit are shown in Table 1. In this table we list the abundance in solar units as well as the column density of the elements. The edge is the approximate location of the strongest absorption edge in or near the RGS band, and τ the optical depth at the edge. The zero column density of carbon is not real but probably a consequence of dust scattering. In contrast, the strong over-abundance of Ne and the under-abundance of iron are reliable and may be signatures that indeed the line of sight contains significant amounts of dust. The oxygen column density of $2.02 \cdot 10^{18}$ cm$^{-2}$ is larger than the value found by Willingale et al.[9] of $1.85 \cdot 10^{18}$ cm$^{-2}$. The present RGS value is more accurate as the edges of the different elements are resolved. The other elemental abundances could not be constrained as accurately as the estimated optical depth at the K-edge is only 0.03 for Mg, 0.02 for Si and less for other elements.

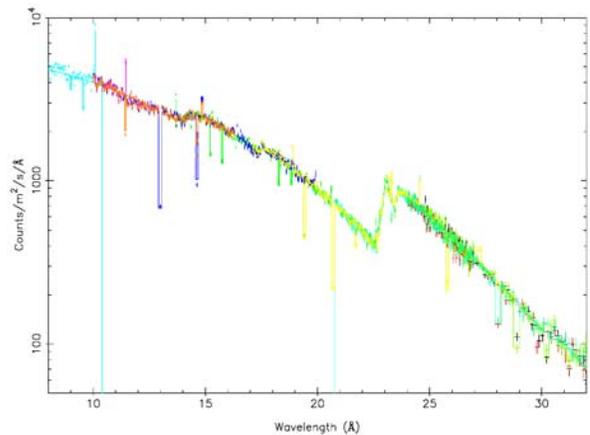

**Figure 5: RGS spectrum of the Crab nebula. Different colours indicate different sets of observations.**

| ELEMENT | ABUNDANCE [SOLAR] | COLUMN [$10^{20}$ CM$^{-2}$] | EDGE [Å] | τ |
|---|---|---|---|---|
| H | 1 | 37.7±0.04 | | |
| C | 0 | - | 43.05 | 0 |
| N | 1.04±0.13 | 0.0031±0.0004 | 30.77 | 0.2 |
| O | 0.93 ±0.007 | 0.0202±0.0002 | 23.05 | 1.01 |
| Ne | 2.51±0.04 | 0.0084±0.0002 | 14.25 | 0.25 |
| Fe | 0.456±0.019 | 0.00060±0.00003 | 17.39 | 0.05 |

**Table 1: Abundances measured by RGS**

Finally, we note that our best-fit continuum normalization is 0.83 with respect to the Kuiper et al.[24] continuum; as our spectrum has the best statistics near 1 keV, this normalisation is a good representation of the measured flux around that energy. An unknown part of this 17 % loss may be due to dust scattering losses, while the remainder could be caused by a small amount of unaccounted pile-up or global cross-calibration corrections.

## 4. RESULTS OF OTHER MISSIONS IN ORBIT

### 4.1. Integral JEM-X

JEM-X (Joint European Monitor for X-rays) is the X-ray monitor aboard INTEGRAL. It consists of two identical coded-mask telescopes, each equipped with an imaging micro strip gas counter. For this analysis we used JEM-X1 data from 30 science windows in revolution 0170. Pointings with off-axis angles up to 4.74 degrees were used. Spectra were extracted with software version OSA4.1. No systematic errors were included. Systematic uncertainties are most important near the lower energy limit. For spectroscopy the useful energy range is approximately 6-30 keV. Due to the uncertain response at low energies absorption columns can in general not be determined from JEM-X observations. The JEM-X instruments are calibrated by Crab observations and 3 of our 30 science windows were also used for those calibrations.

### 4.2. INTEGRAL ISGRI and SPI

The two main instruments aboard INTEGRAL are imager IBIS (Ubertini et al. 2003[47]) and spectrometer SPI (G. Vedrenne et al. 2003[48]). IBIS is a large area (~900 cm$^2$) gamma-ray telescope with two layers of pixellated detectors: ISGRI, made of CdTe and covering the range from 15 keV to 1 MeV, and PICsIT, made of CsI(Tl) and covering the range from 170 keV to 10 MeV. SPI also has a large area (~500 cm$^2$), consisting of 19 HPGe detectors and covering the range from 20 keV to 8 MeV. The INTEGRAL Crab observations were taken during revolutions 0039 (7-10 Feb 2003), 0043 (19-22 Feb 2003), 0044 (22-25 Feb 2003), 0102 (14-17 Aug 2003), 0170 (4-7 Mar 2004), 0239 (27-30 Sep 2004). Spectral extraction for ISGRI was done with OSA 4.2 (released on 15 December 2004). Spectral extraction for SPI and PICsIT was done with pre-OSA 5.0 (version of 14 June 2005). For the SPI spectra we used observations from Revs. 0043, 0044 and 0102 in dithering mode with an offset angle < 6 degrees. The exposure was 388.1 ks. No systematic errors were added and the data were fitted in the useful energy range from 30-1000 keV. SPI seems to be well calibrated, spectral slope is steeper than the Crab canonical model, what is understandable for higher energies when some softening is expected due to the variable contribution from the pulsar that increases with energy. For ISGRI spectra we used observations from Revs. 0043, 0044, 0102, 0170 and 0239 in dithering mode with off-set angle < 6 degrees with a total exposure of 481.9 ks.

### 4.3. Swift XRT

Swift XRT is designed to detect and localize GRB and provides autonomous rapid-response observations and long-term monitoring of their afterglow emission in the X-ray and UV/optical band. The observatory incorporates three primary instruments: the Burst Alert Telescope (BAT[26]), the X-ray telescope (XRT[27]) and the Ultra-Violet/Optical Telescope (UVOT[28]). The XRT is a focusing X-ray telescope operating in the 0.2-10 keV energy band and supports four different read-out modes (IM, PD (PUPD, LRPD), WT, PC) (see Hill et al.[29] for an exhaustive description of XRT observing modes). The XRT effective area for the three XRT observing modes (LRPD, WT and PC) has been calibrated using the Crab nebula and PSR B0540-69 the former for LRPD and WT modes, the latter for PC mode. To calibrate the LRPD ancillary files we consider an on-axis 6742 s exposure of the Crab, resulting in $5.5 \cdot 10^6$ counts in the energy range 0.5-10 keV. The high absorption at low energies does not allow calibrating the spectrum below 0.5 keV. A procedure of ARF optimisation was applied to the LRPD ancillary files generated from on-ground calibrations in order to have a good description of the Crab data with the spectral model parameters reported in the literature. This procedure allows the production of a LRPD ARF file that, when applied to the Crab, reproduces its spectral energy distribution with best-fit parameters consistent with those reported in previous work with other satellites (BeppoSAX: Massaro et al. 2000[30]; RXTE: Pravdo et al. 1997[31]).

### 4.4. Chandra: ACIS

The Chandra spectrum shown here was taken using ACIS-S3 on 2003 January 5. This observation was originally performed to witness the Titan transit of the Crab Nebula. In order to avoid severe event pile-up, the CCD frame time was shortened to 0.3 s and the High-Energy Transmission Grating was inserted: the spectrum shown here is the 0$^{th}$ order spectrum. The entire nebula was not covered fully since the short frame time necessitated the use of a subarray.

However, the missing fraction of the X-ray flux is less than a few percent and is smaller than the systematic errors discussed here. The image and other details of the observational setup are presented in Mori et al. (2004)[32]. The data were re-processed using the latest software package available (CIAO 3.2.1 and CALDB 3.0.3). The pulsar region was excluded because most events in this region had already been discarded on-board due to severe event pile-up. The spectral fitting was performed excluding the 1.5-2.5 keV band where the systematic uncertainty of the effective area is larger than in the rest of the bandpass (Canizares et al. 2005[33]).

### 4.5. RXTE: PCA, HEXTE

The RXTE observation of the Crab on 2000 December 16 has been used to derive 3-240 keV, best-fit parameters for the two-component spectrum. PCU2[34] data covered 3-60 keV, and HEXTE[35] data covered 17-240 keV with a large overlap between the two instruments. Systematic errors of 0.5 % were added to the PCU2 data while none were added to the HEXTE data. The value for the line-of-sight effective hydrogen density was taken from Willingale et al[9] to be $3.45 \cdot 10^{21}$ cm$^{-2}$. After a minor correction to the PCU2 background of -9.04±1.14 percent and ignoring a systematic feature in the PCU2 from 30-35 keV, we fitted the combined PCU2/HEXTE spectrum from the Crab, and found a value for the nebula power law index of 2.146±0.004 (90% errors) and for the weaker pulsed power law index of 1.623±0.034. The relative normalization of HEXTE to PCU2 was 0.984±0.003. The 2-10 keV nebula flux was found to be $2.20 \cdot 10^{-8}$ ergs·cm$^{-2}$s$^{-1}$ and that of the pulsar was $8.3 \cdot 10^{-10}$ ergs·cm$^{-2}$s$^{-1}$. The flux at 1 keV was 11.09 for the nebular emission and 0.187 for the pulsed component. In addition individual fits to PCA and HEXTE have been performed (see Table 2).

## 5. RESULTS OF PAST MISSIONS

### 5.1. ROSAT: PSPC

The Crab was observed by ROSAT[49] several times during the pointed programme, both with the *Position Sensitive Proportional Counter* (PSPC) and the *High Resolution Imager* (HRI). It was not used for tuning spectral calibrations. The high count rate (~ 750 cts/s) causes a reduction of the gain in the PSPC of about 5 %, which distorts the spectrum. The command *PROCESS/CT* in the EXSAS package allows redoing the standard pipeline processing from raw to calibrated events with a user interface to calibration files. For the spectral analysis presented here the observation *500065p* (March 1991, ROSAT day 277, 9.1 ks) has been taken once with standard gain correction (events files extracted from the archive) and once processed from raw telemetry events with a reduced gain (138.35 compared to 145.63; this does not fully reflect the energy dependence involved). No additional spatial gain map had been used. Only events within the inner PSPC ring have been selected (<18'), standard dead time correction was applied but no vignetting correction (as the source was on-axis and events in the PSF wings would otherwise obtain an over-correction), high background periods have been discarded (leaving 6 ks and 4424318 events in the 0.1-2.4 keV band). No further background subtraction was required due to the source brightness and the very low intrinsic PSPC background.

### 5.2. EXOSAT

EXOSAT spectra obtained with the Medium Energy (ME) instrument and the Gas Scintillation Proportional Counter (GSPC) together with the response matrices were extracted from the HEASARC archive. The seven ME spectra from observations between October 1983 and March 1986 were analysed in the 1-20 keV energy band. From the GSPC observations five spectra are available which we fit in the 2-15 keV band. For the EXOSAT calibration the Crab was used as standard with "canonical" values of $3 \cdot 10^{21}$ cm$^{-2}$ for the $N_H$ and 2.08 for the photon index using XSPEC with the *wabs* absorption model.

### 5.3. BeppoSAX : LECS, MECS, HGSPC, PDS

BeppoSAX spectra for the imaging spectrometers (LECS, Parmar et al. 1997[36]; MECS, Boella et al. 1997[37]) were extracted from circular regions of 8 arcmin. Background spectra were extracted from blank sky event lists, from the same region in detector coordinates as the nebula. Data from the non-imaging spectrometers HPGSPC (Manzo et al. 1997[38]) and PDS (Frontera et al. 1997[39]) correspond to the whole fields-of-view of 1°·1°, and 1.3°·1.3°, respectively. Accurate background subtraction in the PDS is guaranteed by a rocking system, which monitor the source and two regions 3.5° off-side every 90 s. Background modelling in the latter is achieved through on-ground estimates, based on environmental housekeeping parameters. A detailed description of the BeppoSAX instrument calibration and data reduction procedures is available in Fiore et al. (1997)[40]. The data presented in this paper refers to an observation taken on 1999 September 29 with the exception of the MECS spectrum, which has been accumulated using all the Crab observations performed by BeppoSAX. The most updated standard response matrices as provided by the ASI Science Data Centre have been employed

### 5.4. ASCA: GIS2, GIS3

ASCA carried a pair of Solid State Spectrometers (SIS) and a pair of Gas Imaging Spectrometer (GIS; Ohashi et al. 1996)[41] cameras. The Crab cannot be observed by the former due to heavy pile-up. We present in this paper spectra taken by the latter on 1994 September 28. Data reduction followed standard procedures. No background subtraction was performed, as the Crab spot in the GIS cameras completely filled their field-of-view. Such a correction is estimated to be negligible. Dead-time correction is instead important for such bright sources, and was performed via the task *deadtime* in the *Lheasoft* package. Effective areas appropriate for each camera have been generated with the Lheasoft tool *ascaarf*, on the basis of standard redistribution matrices distributed by the HEASARC ASCA GOF (version 4).

### 5.5. Ginga

The Ginga spectrum has been extracted from the GINGA archives at Data ARchive and Transmission System (DARTS) at ISAS, Japan (see, http://www.darts.isas.ac.jp/). The observation was made on 1987 March 27 from 05:39 to 05:48. Analysis was made with ISAS GINGA archival analysis software package (http://www.darts.isas.ac.jp/astro/ginga/). Background was subtracted, and the data have been corrected for dead time and vignetting.

### 5.6. Mir-HEXE

The HEXE instrument aboard the Kvant module of the Mir space station consisted of four phoswich detectors covering the energy range 20-200 keV with an effective area of 750 $cm^2$ (Reppin et al. 1985[42]). The detector energy response matrices were calibrated in space assuming a Crab power law slope of 2.08 (Toor & Seward 1974[6]); a value consistent with the observed data >80 keV when using responses data predicted from ground calibrations without further corrections (Kretschmar, 1991[43]). Using these matrices the best fit results for the total Crab spectrum were A = 8.89(7) and Γ = 2.08(2).

### 5.7. Einstein

The Einstein Observatory[44] observed the Crab nebula with both imaging instruments, the HRI and the IPC. For the IPC, the nebula produced a count rate of approximately 1300 cts s$^{-1}$, which severely saturated the telemetry capability of the detector. Furthermore the IPC was not calibrated at such high rates and therefore the IPC data cannot be reliably analysed. The HRI with its lower sensitivity (20 $cm^2$ at 0.28 keV and 5 $cm^2$ at 2 keV) was used to observe the Crab nebula twice[45]. In the energy band from 0.1 to 4.5 keV, the HRI count rate of the nebula was approximately 120 cts s$^{-1}$. Since the HRI had no spectral resolution, the flux and the luminosity of the nebula can only be given by assuming a source spectrum. Harnden and Seward used a power-law spectrum with an energy index of 1.1 and found for the 0.1 to 4.5 keV band an absorption-corrected flux from the entire nebula of 7.8·10$^{-8}$ erg cm$^{-2}$s$^{-1}$ assuming $N_H=3\cdot10^{21}$ cm$^{-2}$. Assuming a distance to the nebula of 2.0 kpc, this corresponds to a luminosity of $3.7\cdot10^{37}$ erg s$^{-1}$. The errors of this flux and luminosity determination are dominated by systematic uncertainties of the absolute calibration of the HRI, which is of the order of ± 15%.

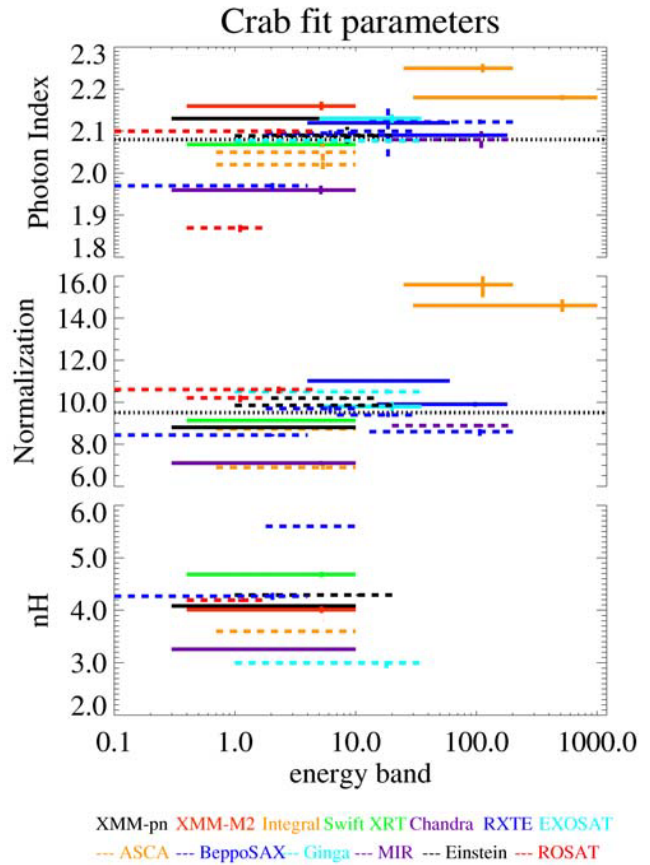

Figure 6: Spectral parameters of the Crab for various X-ray missions. The horizontal bars represent the instrument's operational energy range. $N_H$ is given in $10^{21}$ cm$^{-2}$ and normalization in photons keV$^{-1}$cm$^{-2}$s$^{-1}$ at 1 keV.

## 6. COMPARISON OF THE VARIOUS DATA

In order to get a handle of the huge amount of different data we provide the following information: first, we summarize the fit parameters of the individual instruments to allow direct comparison of the absolute values from each mission; second, we perform a joint fit to all datasets only allowing for a normalisation constant for each instrument to vary; and thirdly, we provide fitting parameters of joint fits for different energy bands (0.1-2 keV, 2-10 keV and 10-50 keV and 50-1000 keV).

### 6.1. Individual fits

Each dataset has been fit with an absorbed power law, once using abundances of Wilms et al. and cross-sections of Balucinska-Church & McCammon and once using abundances of Wilms et al. and cross-sections of Verner et al. No systematic errors were included for this comparison. In Table 2 below we show the results of those fits, indicating which instruments have originally been calibrated on the Crab (pink). Figure 6 gives an overview of the spectral parameters of the Crab derived by each instrument, indicating the relevant instrumental energy ranges.

| | | PHABS [ABUND WILM XSECT BCMC] | | | PHABS [ABUND WILM XSECT VERN] | | |
|---|---|---|---|---|---|---|---|
| INSTRUMENT | ENERGY | $N_H[10^{21}$ CM$^{-2}]$ | $\Gamma$ | N | $N_H[10^{21}$ CM$^{-2}]$ | $\Gamma$ | N |
| XMM pn | 0.3-10 | 3.81(3) | 2.125(4) | 8.86(2) | 4.08(2) | 2.130(3) | 8.80(4) |
| XMM MOS2 | 0.4-10 | 3.76(6) | 2.15(1) | NA | 4.01(6) | 2.16(1) | NA |
| XMM RGS | NA | NA | NA | NA | NA | NA | NA |
| Chandra | 0.5-1.5 2.5-10 | 3.05(2) | 1.95(3) | 7.04(3) | 3.26(3) | 1.96(3) | 7.10(3) |
| Swift | 0.4-10 | 4.37(5) | 2.057(5) | 9.04(2) | 4.68(5) | 2.068(5) | 9.14(2) |
| RXTE HEXTE | 15-180 | 3.8 (f) | 2.09(2) | 9.9(1) | 4.0 (f) | 2.090(2) | 9.9(1) |
| RXTE PCA | 4-60 | 3.8 (f) | 2.120(2) | 11.02(4) | 4.0 (f) | 2.120(2) | 11.02(4) |
| INT. JEM_X | 5-35 | 3.8 (f) | 2.136(8) | 9.8(2) | 4.0 (f) | 2.136(8) | 9.8(2) |
| INT. ISGRI | 25-200 | 3.8 (f) | 2.253(3) | 15.4(5) | 4.0 (f) | 2.252(2) | 15.47(2) |
| INT. SPI | 30-1000 | 3.8 (f) | 2.203(3) | 15.9(1) | 4.0 (f) | 2.203(3) | 15.9(1) |
| BSAX/LECS | 0.1-4.0 | 4.00(3) | 1.973(6) | 8.33(4) | 4.27 (7) | 1.977 (6) | 8.43 (5) |
| BSAX/MECS | 1.8-10.5 | 5.4(4) | 2.096(8) | 9.7(2) | 5.6(4) | 2.092(8) | 9.7(2) |
| BSAX/HPGSPC | 7.0-30.0 | 3.8 (f) | 2.10(6) | 9.4(1s) | 4.0 (f) | 2.10(6) | 9.4(1) |
| BSAX/PDS | 13.0-200.0 | 3.8 (f) | 2.126(2) | 8.83 (7) | 4.0 (f) | 2.126(4) | 8.84 (7) |
| ASCA/GIS2 | 0.7-10.0 | 3.4(2) | 2.02(1) | 6.9(1) | 3.6 (2) | 2.02 (1) | 6.9 (1) |
| ASCA/GIS3 | 0.7-10.0 | 3.5(2) | 2.05(1) | 8.78(1) | 3.6 (2) | 2.05 (2) | 8.74 (2) |
| ROSAT | 0.4-1.8 | 3.8(2) | 1.83(3) | 9.9(2) | 4.19(3) | 1.87(1) | 10.2(2) |
| EXOSAT ME | 1-20 | 4.15(2) | 2.088(3) | 9.84(7) | 4.29 (2) | 2.089(2) | 9.85(2) |
| EXOSAT GSPC | 2-15 | 3.8 (f) | 2.09(3) | 10.14(7) | 4.0 (f) | 2.09(2) | 10.19(5) |
| GINGA | 1-35 | 2.9(2) | 2.077(4) | 10.5(1) | 3.0(1) | 2.077(4) | 10.5(1) |
| EINSTEIN | 0.1-4.5 | 3.0 | 2.1 | NA | NA | NA | NA |
| MIR XEXE | 20-200 | NA | 2.08(2) | 8.89(7) | NA | NA | NA |

**Table 2:**
**Parameters of the fits to individual instruments.**
**The normalization *N* is given in photons keV$^{-1}$cm$^{-2}$s$^{-1}$ at 1 keV. Instruments calibrated on the Crab are marked in pink. The errors are 90 % confidence.**

Note that our parameters for $N_H$ differ in absolute terms from previous values since we use a different absorption model and not the obsolete *wabs* model. Fitting the EPIC-pn data with the *wabs* model results in an $N_H$ of $3.0 \cdot 10^{21}$ cm$^{-2}$ with respect to $3.81 \cdot 10^{21}$ cm$^{-2}$ and $4.08 \cdot 10^{21}$ cm$^{-2}$ with the new models. Note that the comparison of the 1 keV flux can be misleading for the high energy instruments since tiny differences in the power-law index are hugely magnified. Currently work is underway to determine the "best-fit flux" in several bands and to compare the flux for each instrument in the appropriate band to the "best-fit flux" in that band. This ratio will give a more appropriate flux comparison.

The EPIC cameras of XMM-Newton suggest a lower $N_H$ than EXOSAT, Swift and BeppoSax and higher than Ginga, ASCA and Chandra. However EXOSAT, Swift and BeppoSax have been calibrated to the Crab and it may well be that the EPIC with their immense throughput can refine the $N_H$ value of the Crab.

### 6.2. Simultaneous fits

In order to compare the individual instruments we fitted all data simultaneously in various energy ranges (see Tab. 3) using a simple absorbed power law model (*phabs\*powerlaw*) with abundances set to Wilms et al and cross sections to Verner et al. Allowing for normalization constants for all spectra fixing the EPIC-pn constant to 1 and linking the individual parameters we derive for a fit from 0.1-1000 keV the following spectral parameters of the Crab: $N_H=4.5 \cdot 10^{21}$ cm$^{-2}$, $\Gamma=2.08$ and N=8.97. Given the good statistic the result of the sum of all calibration deficits of course produces an unacceptable reduced chi$^2$ value. We therefore report no errors in the following.

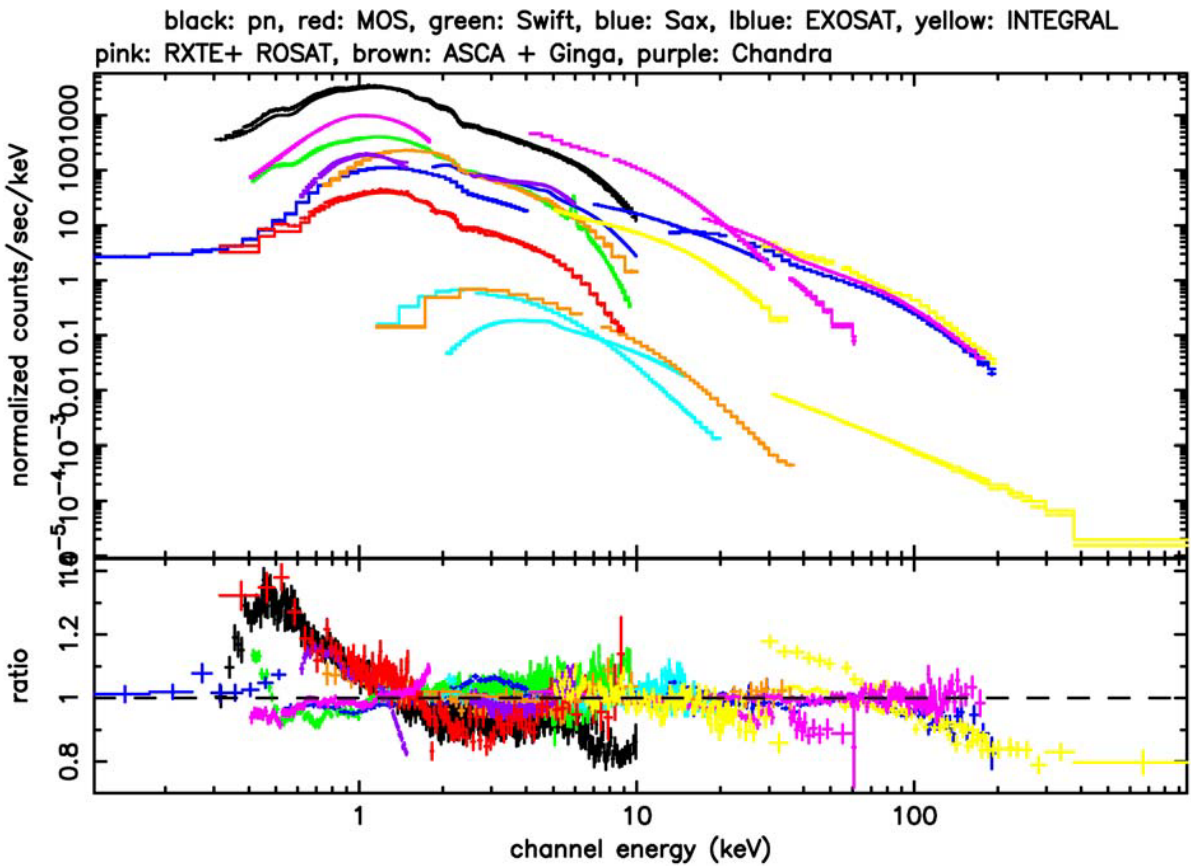

**Figure 7: Joint fit to all data in the energy range 0.1-1000 keV**

The interesting features in Figure 7 are actually the deviation form the canonical or the overall Crab fit that can be seen in the residuals in the lower panel: ROSAT, LECS and Swift XRT seem to agree at energies below 1 keV, while the EPIC instruments, Chandra, ASCA seem to see more flux. The EPIC-pn camera shows a flux deficit above 7 keV, which is most probably related to an imperfect treatment of the effective area of the mirror module, that is in front of the EPIC-pn camera. This is currently under study and may be updated soon with a new X-ray telescope effective area Current Calibration File. ASCA seems to agree with EPIC, but unfortunately covers not the full EPIC energy range. INTEGRAL SPI shows a steeper photon index which may be related to the intrinsic change of the Crab spectrum for higher energies.

Table 3: Joint fit parameters in different energy ranges

| ENERGY RANGE IN KEV | $N_H(H)$ | $\Gamma$ | N |
|---|---|---|---|
| 0.2-2 | 4.07 | 2.02 | 8.95 |
| 2-10 | 4.5 (f) | 2.07 | 8.26 |
| 10-50 | 4.5 (f) | 2.12 | 9.42 |
| 50-1000 | 4.5 (f) | 2.17 | 10.74 |
| 0.1-1000 | 4.5 | 2.08 | 8.97 |

## 7. SUMMARY AND CONCLUSIONS

Most of the instruments that have their main sensitivity range above 2 keV cannot be used efficiently to constrain the absorption towards the Crab and have been calibrated to fit canonical Crab parameters. We believe that this fact drives us most probably in the direction to assume that XMM-Newton may revise the X-ray absorption of the Crab, given EPIC's very high effective area in the low-energy regime. The photon index and normalization provided by EPIC agree within the errors of the Toor and Seward values taking also into account the systematic errors of EPIC on the photon index (0.05). The normalisation of EPIC-pn however seems quite low in comparison with all other instruments and needs to be checked by careful treatment of the EPIC-MOS data in order to derive also from those data an additional EPIC normalization. Further work needs to be carried out to combine the RGS results of the under-abundances with the EPIC results. This has been done already in a preliminary analysis but shows only some changes at the fine structure of the EPIC low-energy excess. The general trend of EPIC giving a lower column density is still present.

XMM-Newton may soon provide the best calibrated spectrum of the Crab. Given the extensive calibration campaigns at the PANTER test facility and the Orsay LURE synchrotron XMM-Newton provides a very good pre-launch calibration and the cameras have later NOT been calibrated on the Crab. The current discrepancy between EPIC-pn and EPIC-MOS in photon index is comparable to the scatter among all instruments. Resolving the pn-MOS discrepancy will provide a precise and accurate (and modern!) spectrum of the Crab with no assumptions about the spectrum built into the calibration. The modern spectrum will contain high-quality measurements of the absorption, in contrast to previous efforts.

The range of measurements that we report is comparable to the uncertainty reported by Toor and Seward many years ago (we tend to remember their answer, but not their errors). The prospects for reducing our systematic errors below this level are good.

For energies above 30 keV the Crab is definitely difficult to use as calibration source. Since its spectrum is no longer a single power law no consensus obtains on the proper model of the spectrum which should be used for calibration tests. Different results may well be caused by the more complex character of these instruments which are generally difficult to calibrate and have a large intrinsic background that increases with energy

Given the fact that the Crab will be possibly too bright and too extended for future X-ray missions one may think about establishing a set of standard calibration sources for the X-ray regime.[50]